\documentclass[a4paper,11pt]{article}
\usepackage{pos}
\usepackage{float,physics}
\usepackage{hyperref}
\usepackage{multicol}

\title{Extracting Yang-Mills topological structures with adjoint modes}

\author*[a]{I.~Soler}
\author[a]{G.~Bergner}
\author[b]{A.~González-Arroyo}

\affiliation[a]{University of Jena, Institute for Theoretical Physics,\\
Max-Wien-Platz 1, D-07743 Jena, Germany}

\affiliation[b]{
Departamento de Física Teórica C-15 and \\
Instituto de Física Téorica UAM-CSIC, \\
Universidad Autónoma de Madrid, Cantoblanco, E-28049 Madrid, Spain}

\emailAdd{ivan.soler.calero@uni-jena.de}
\emailAdd{georg.bergner@uni-jena.de}
\emailAdd{antonio.gonzalez-arroyo@uam.es}

\abstract{We report on how adjoint zero modes can be used to filter out the topological structures of gauge configurations from the UV fluctuations. We will use the Adjoint Filtering Method (AFM) which relies on the existence of a particular Supersymmetric Zero Mode (SZM) that follows closely the (anti)self-dual part of the action density. In contrast, it is not guaranteed that summing over the lowest lying modes in the fundamental representation reproduces the topological density. We present preliminary qualitative results on smooth, heated, and Monte-Carlo generated configurations with non-trivial content of fractional instantons. The method is capable of distinguishing the underlying topological structures without significantly modifying the gauge field as opposed to smoothing techniques. The method looks promising as a tool to investigate and extend recent studies based on semiclassical methods. 
}

\FullConference{The 40th International Symposium on Lattice Field Theory (Lattice 2023)\\
July 31st - August 4th, 2023\\
Fermi National Accelerator Laboratory\\}

\begin{document}
\maketitle

\section{Introduction}
The properties of the non-peturbative vacuum of gauge theories remain to a large extent still unknown. Promising approaches are based on semiclassical analysis, which try to identify the dominant long range topological contributions. Progress has been made in particular for supersymmetric gauge theories with this approach. Early attempts have been based on instanton gas pictures and a more recent proposals relies on a factional instanton liquid model \cite{Gonzalez-Arroyo:2023kqv}. It is important to test these approaches in a qualitative and quantitative way.
Lattice simulations provide in principle a unique way to investigate non-perturbative vacuum properties of gauge theories. However, for this study it is essential to filter the long range contributions from the ultraviolet (UV) quantum fluctuations.

In order to filter out this UV noise, several smoothing techniques, like the Gradient Flow (GF) or cooling have been used. A disadvantage of these methods is the significant modification of the underlying gauge configuration. This has motivated the search for alternative methods based on eigenmodes of the Dirac operator.
The idea of filtering gauge configurations using fermionic zero modes seems very appealing, as they are naturally very smooth long range quantities. A natural connection between zero modes and topology is given by the celebrated Atiyah-Singer index theorem. For fundamental fermions, there is, however, no precise connection of the lowest modes and the local action or topological charge density of the underlying gauge configuration. For the Dirac operator in the adjoint representation, it was noticed in \cite{Gonzalez-Arroyo:2005fzm} that there is such a precise connection at least for smooth configurations. Based on these findings the Adjoint Filtering Method (AFM) has been proposed and tested for some smooth configurations \cite{GarciaPerez:2011tx}.

In this short proceedings we collect the main ideas of the AFM. We will apply the method first to smooth and heated configuration with a given of fractional insanton background. Then we will move to Monte-Carlo generated configurations.

\section{Adjoint Filtering Method}
Fermions in the adjoint representation can be related to the gauge field by supersymmetry transformations. In particular, in \cite{S.Chadha:1977} the existence of a particular fermionic zero mode was pointed out, which we will refer as the Supersymmetric Zero Mode (SZM). It is given by:
\begin{align}
\psi(V,x)=\frac{1}{8} F_{\mu\nu}[{\gamma_\mu},{\gamma_\nu}]V ,
\label{SZM}
\end{align}
where $V$ is a constant four spinor. There are two SZM distinguished by their chirality $\psi_+$ and $\psi_-$.  Choosing $V=(1,0,0,0)$ for example, we get the positive chirality one
\begin{align}
\psi_+=
\begin{pmatrix}
(E_3+B_3)\\
(E_1+B_1)/2+i(E_2+B_2)/2\\
0\\
0
\end{pmatrix},
\label{SZM:real}
\end{align}
The connection of this zero mode to the stress energy tensor is obvious from Eq.~\eqref{SZM}, the modulus squared of this mode trivially reproduces the self-dual part of the gauge action density, while the oposite chirality reproduces the anti-self dual part
\begin{align}
|\psi_\pm|^2=\frac12 (F_{\mu\nu}\pm \tilde{F}_{\mu\nu})^2.
\label{SZM:density}
\end{align}
Therefore the density of this mode provides a method to filter out the UV noise of gauge configurations and obtain a filtered topological charge density. Indeed if one defines
\begin{align}
    q_\psi(x)\equiv|\psi_+(x)|^2-|\psi_-(x)|^2, \quad q_A(x)\equiv\frac{1}{32\pi^2}\epsilon_{\mu\nu\rho\sigma}\text{tr}[F_{\mu\nu}\tilde{F}_{\mu\nu}],
\end{align}
by using Eq.~\eqref{SZM:density} one can see that $q_\psi(x)\propto q_A(x)$.

Notice also that the first component of the SZM Eq.~\eqref{SZM:real} is purely real, which makes it distinguishable from the rest of the zero modes. We implemented such a condition through the supersymmetric operator
\begin{align}
\mathcal{O}^\pm=P_0P_\pm(\gamma_5 D^A_{ov})^2P_\pm P_0,
\end{align}
where $ D^A_{ov}$ is the Overlap operator in the adjoint representation, $P_\pm=\frac{1}{2}(1\pm \gamma_5)$ projects to the different chirality sectors and $P_0$ to the reality condition for the first component. The eigenvector of the lowest eigenmode of $\mathcal{O}$ is in general an optimization of the three conditions corresponding to the properties of the SZM:
\begin{multicols}{3}
\begin{itemize}
	\item Zero mode of $ D^A_{ov}$.
	\item Chirality condition.
	\item Reality condition.
\end{itemize}
\end{multicols}
We have found that in practice for noisy configurations not only a single mode has to be considered. Instead one needs to sum several lowest modes of the operators.
As a first non-trivial check we will start by applying the method to configurations with a given semiclassical background and then we will move to Monte-Carlo generated ones.
\section{Test configurations}
The non-trivial background of the test configurations has been produced starting from configurations with non-trivial twist and a fractional instanton. These topological objects are of physical interest since they are the basic contributions in a fractional instanton liquid model \cite{Gonzalez-Arroyo:2023kqv}.
Smooth fractional instantons can be obtained by using cooling techniques on a lattice with twisted boundary conditions, as fractional instantons are the ground state in this case. By using time reversal, which transforms instantons into anti-instantons, and gluing several of these configurations one obtains the desired topological content on a larger lattice. Afterwards, one can introduce noise by heating the configuration with standard heat bath steps.

\subsection{Instanton/Anti-instanton pair: $SU(2)$, $\ V=16\times8^3$, $\  Q=0$}
\begin{figure}[h!]
    \centering
    \includegraphics[width=0.30\linewidth]{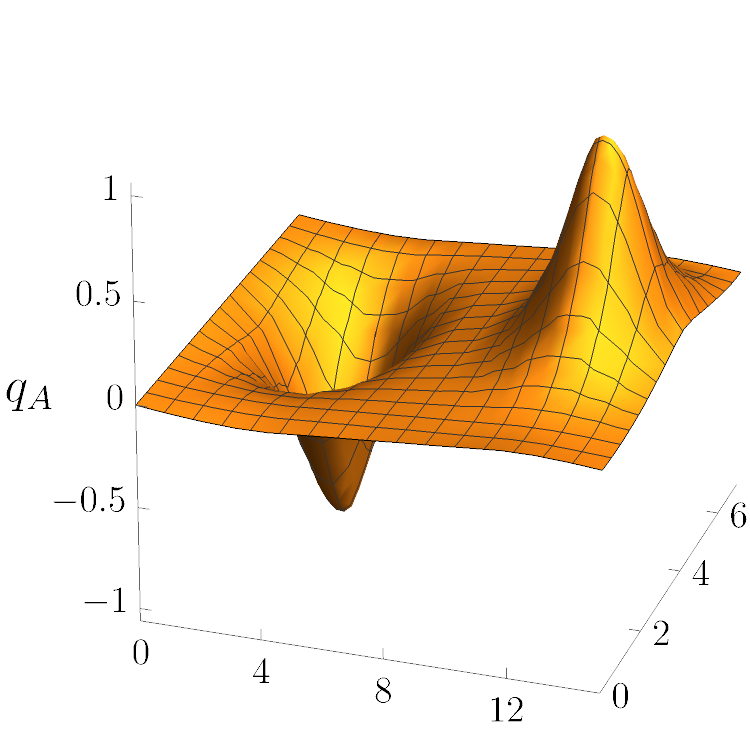}
    \hspace{1cm}
    \includegraphics[width=0.30\linewidth]{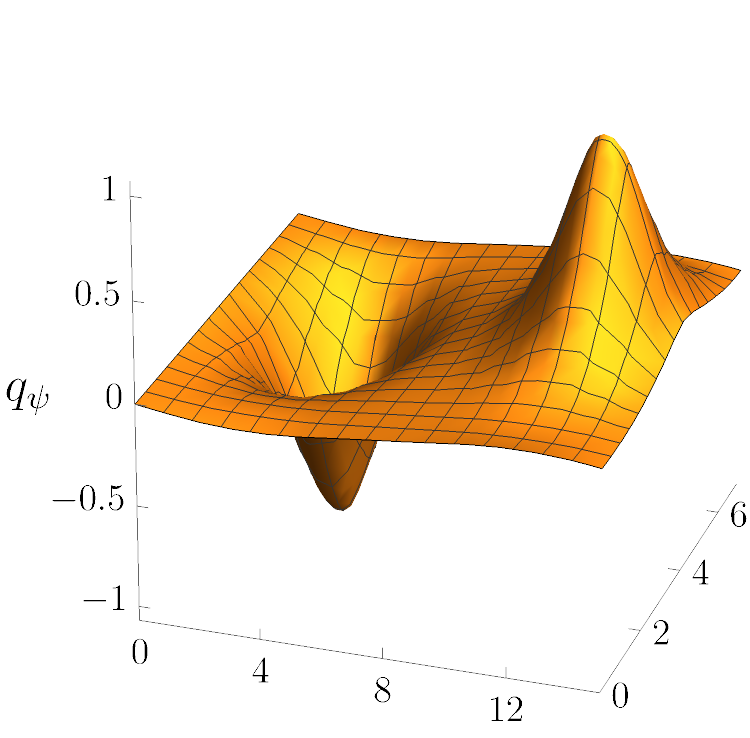}
        \caption{AFM applied to an instanton/anti-instanton configuration with $Q=0$. The two figures compare gauge field topological density to AFM.}
    \label{Q0_AFM} 
\end{figure}

An important example are instanton and anti-instanton pairs. 
These configurations are particularly difficult as they do not correspond to minima of the classical action. This means cooling methods or GF will annihilate the pairs. Furthermore, the fermionic modes associated to these configurations are lifted to quasi-zero modes. Nevertheless, we observe a gap on the spectrum of the Overlap and the $\mathcal{O}$ operator Tab.~\ref{tab:AFM Q0}. Remarkably, the SZM can be identified and its density reproduces the topological density Fig.~\ref{Q0_AFM}.

\begin{table} [h!]
\centering
    \begin{tabular}{|c|c|c|c|}
        \hline
        $\mathcal{O}_-$ & $\mathcal{O}_+$ & $P_-D_{ov}P_-$ & $P_+D_{ov}P_+$ \\
        \hline
        $\lambda_1= 1.17 * 10^{-4}$&$\lambda_1= 1.17 * 10^{-4}$ &$\lambda_{1,2} = 1.03 *10^{-4}$ &$\lambda_{1,2}=  1.03 *10^{-4}$ \\
        \hline
        $\lambda_2= 1.75 * 10^{-2}$&$\lambda_2= 1.75 * 10^{-2}$&$\lambda_{3,4} = 1.51 * 10^{-2}$ &$\lambda_{3,4}= 1.51 * 10^{-2}$ \\
        \hline
        $\lambda_3= 2.05* 10^{-2}$&$\lambda_3=  2.05* 10^{-2}$ &$\lambda_{5,6} = 2.47 *10^{-2}$&$\lambda_{5,6}=2.47 *10^{-2}$\\
        \hline
    \end{tabular}
    \caption{Lowest eigenvalues of the $\mathcal{O}_\pm$ and $P_\pm D_{ov}P_\pm$ operators for a smooth $V=16\times8^3$, $Q=0$ configuration.}
\label{tab:AFM Q0}
\end{table}

\begin{figure}[h!]
    \centering
    \includegraphics[width=0.40\linewidth]{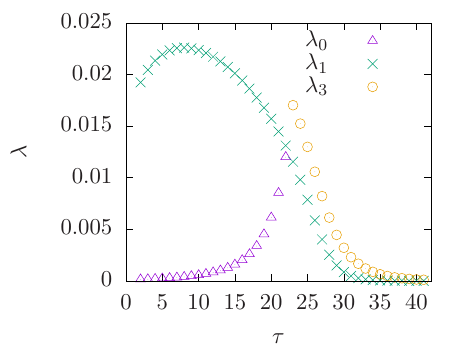}
    \includegraphics[width=0.40\linewidth]{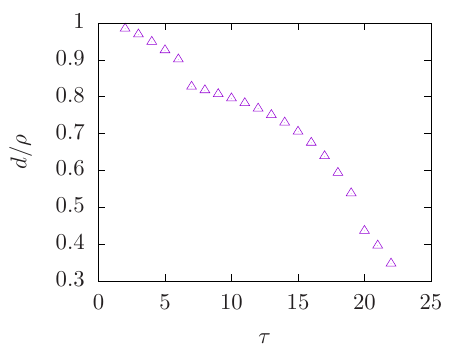}
        \caption{Left: lowest modes of the $\mathcal{O_+}$ operator obtained by flowing the $Q=0$ configuration. Right: evolution of the distance over the size of the instantons $d/\rho$ during flow }
\label{fig:Q0_AFM_GF}
\end{figure}

We can go one step further and use the GF to modify the configurations inducing an annihilation process of the pair. In this way, we can study how the AFM performs as the distance $d$ over the size of the instantons $\rho$ decreases Fig.~\ref{fig:Q0_AFM_GF}. We can see that the lowest eigenvalue raises with increasing flow time and the pair gets closer together. At a certain flow time there is a level crossing of the eigenvalues of $\mathcal{O}$ and what becomes the lowest mode does not represent the topological density anymore. Therefore according to the AFM there is a clear point of annihilation and short distant pairs are still resolved. This also shows the general shortcoming of the GF as it can even destroy the structures one is trying to filter from the noise.

\subsection{Heated configuration: $SU(3)$, $\ V=40^2\times6^2$, $\  Q=0$}
\begin{figure}[h!]
	\centering
	\includegraphics[width=0.30\linewidth]{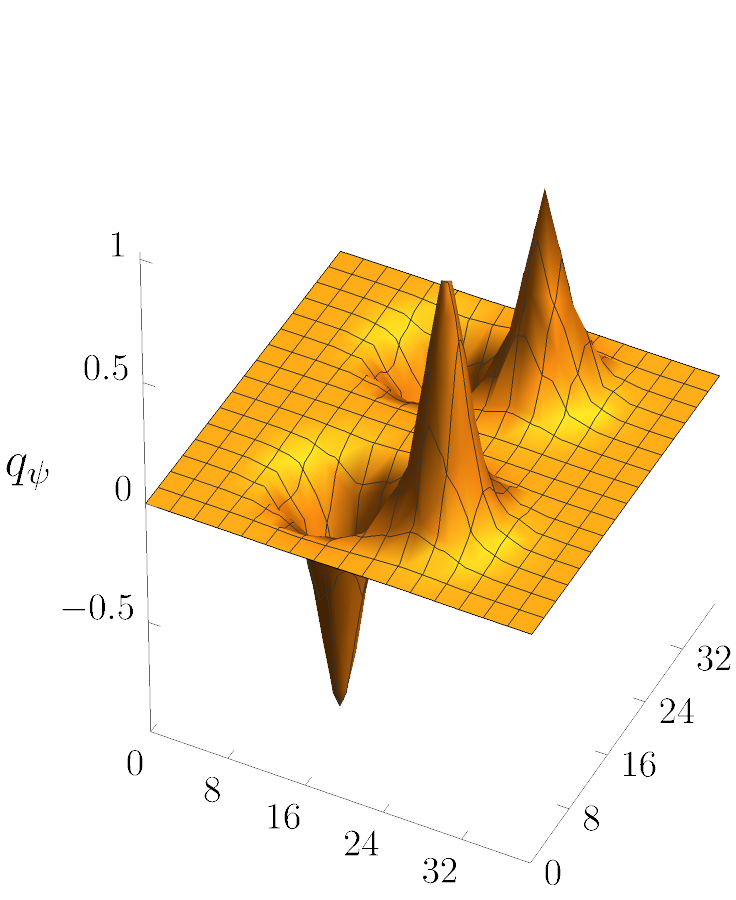}
	\hspace{0.5cm}
	\includegraphics[width=0.30\linewidth]{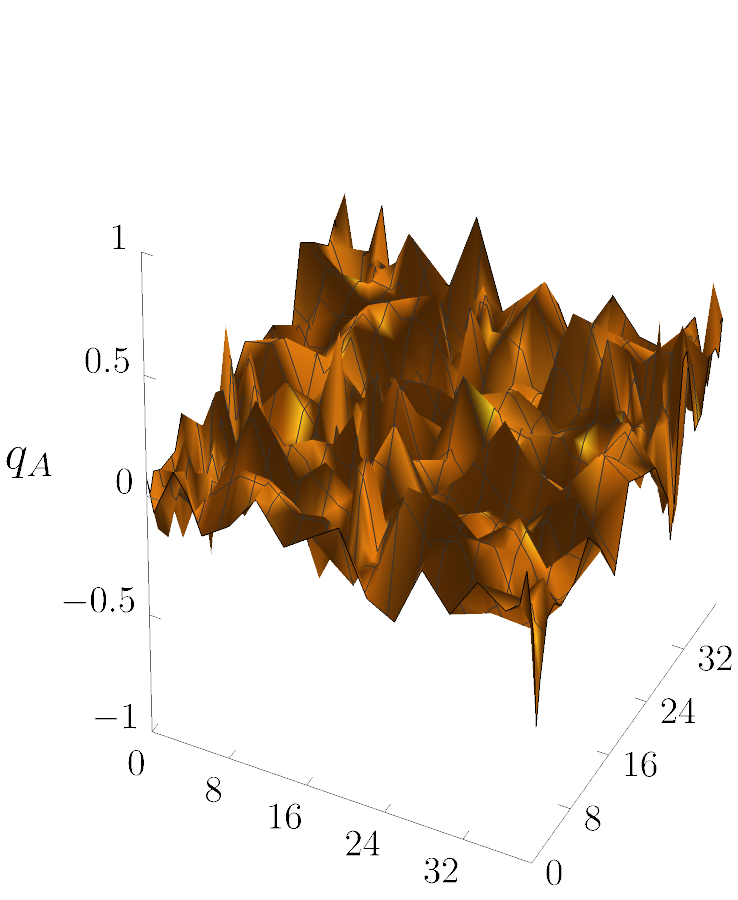}
	\hspace{0.5cm}
	\includegraphics[width=0.30\linewidth]{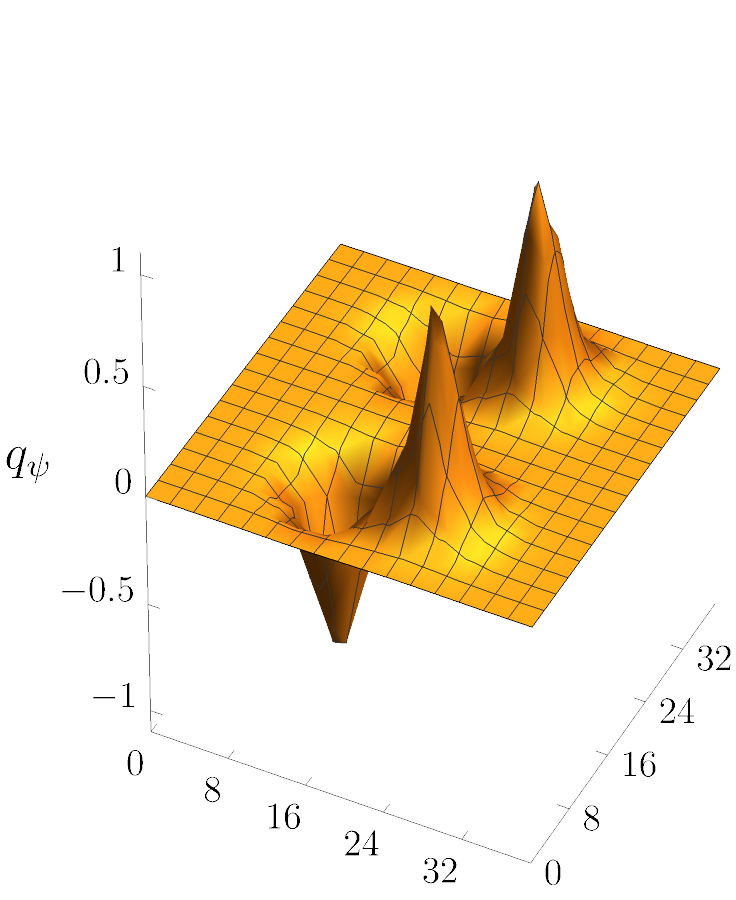}
	\caption{\textbf{Left plot}: SZM zero mode density on smooth configuration before heating. \textbf{Middle plot:} Topological density of the heated configuration. \textbf{Right plot:} Density of the SZM of the heated configuration.}
	\label{fig:40x6}
\end{figure}

In the next step we study the effect of noise on the smooth configurations. We consider a test configuration based on $SU(3)$ pure Yang-Mills on a $V=40^2\times6^2$ lattice. Initially the configuration presents two fractional instanton/anti-instanton pairs. The spectrum of the $\mathcal{O_{\pm}}$ operator on the initial configuration Tab.~\ref{tab:40x6_smooth_spectrum} shows two modes clearly gaped from the rest of the spectrum and the density of the lowest mode for each chirality matches correctly the gauge density, Fig.~\ref{fig:40x6} (left).
\begin{table}[h!]
	\centering
	\begin{tabular}{|c|c|}
		\hline
		$\mathcal{O}_-$ & $\mathcal{O}_+$ \\
		\hline
		$\lambda_1= 4.79 * 10^{-5}$ & $\lambda_1= 4.79 * 10^{-5}$  \\
		\hline
		$\lambda_2= 4.95 * 10^{-5}$ & $\lambda_2= 4.95 * 10^{-5}$\\
		\hline
		$\lambda_3= 2.16* 10^{-2}$&$\lambda_3=  2.16* 10^{-2}$ \\
		\hline
		$\lambda_4= 2.16* 10^{-2} $&$\lambda_4= 2.16* 10^{-2}$ \\
		\hline
	\end{tabular}
    \hspace{1cm}
    \begin{tabular}{|c|c|}
		\hline
		$\mathcal{O}_-$ & $\mathcal{O}_+$ \\
		\hline
		$\lambda_1= 3.32 * 10^{-2}$ & $\lambda_1= 3.38 * 10^{-2}$ \\
		\hline
		$\lambda_2= 3.6* 10^{-2}$ & $\lambda_2= 3.56 * 10^{-2}$ \\
		\hline
		$\lambda_3= 8.5* 10^{-2}$&$\lambda_3=  9.26* 10^{-2}$ \\
		\hline
		$\lambda_4= 8.7* 10^{-2} $&$\lambda_4= 9.5* 10^{-2}$ \\
		\hline
	\end{tabular}
	\caption{Lowest eigenvalues of $\mathcal{O}_\pm$ before and after applying heat bath.}
	\label{tab:40x6_smooth_spectrum}
\end{table}


After several heat bath steps the structures can not be identified from the gauge field density, Fig.~\ref{fig:40x6} (middle). The spectrum of the $\mathcal{O}$ operator shows a smaller gap between the lowest modes and the next states, Tab.~\ref{tab:40x6_smooth_spectrum}. The lowest mode only captures one of the fractional instantons. However, the sum of the lowest two modes on each sector correctly reproduces the underlying gauge density before heating , Fig.~\ref{fig:40x6} (right). 

\section{Monte-Carlo configurations} 
For Monte-Carlo configurations we chose a $V=32\times4^3$ lattice with twisted boundary conditions, standard $SU(2)$ Yang-Mills Wilson action at $\beta=2.44$. 
It turned out that a small GF (flow time $\tau=0.5$) before applying the AFM leads to significant improvements. A detailed analysis of the optimal flow time will be presented in a future publication.  We will first comment on the tuning of the parameters and then show some qualitative results.

\begin{figure}[h!]
	\centering
	\includegraphics[width=0.40\linewidth]{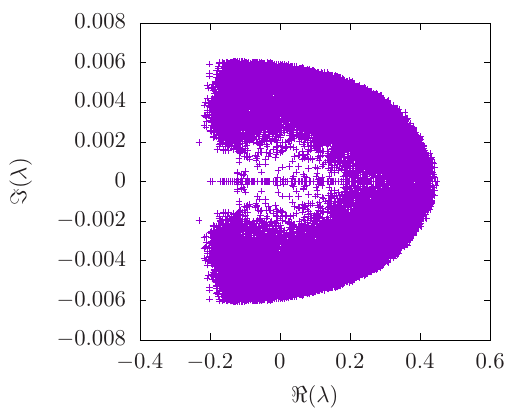}
	\includegraphics[width=0.40\linewidth]{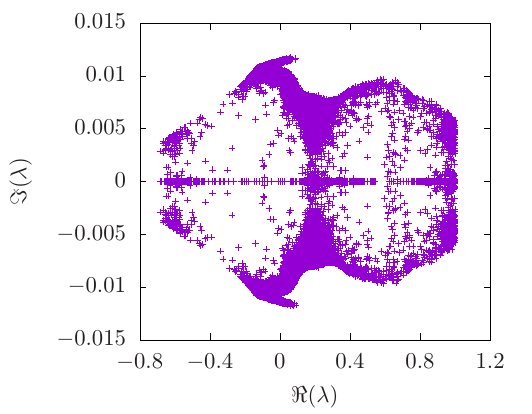}
	\caption{Lowest part of the spectrum of the DW operator at flow time $\tau=0, 0.5$ for 50 configurations.}
	\label{wilson_mc}
\end{figure}

The mass parameter of the Dirac-Wilson kernel in the overlap operator needs to be tuned properly such that doublers are driven towards $\lambda_{ov}=2$ while the physical zero modes are mapped close to the origin. 
The improvement of the small initial GF smoothing is clearly seen from the spectrum of the Dirac-Wilson operator since one obtains a much better separation between the doublers and the rest of the spectrum, Fig.~\ref{wilson_mc}.

\begin{figure}[h!]
	\centering
	\includegraphics[width=0.45\linewidth]{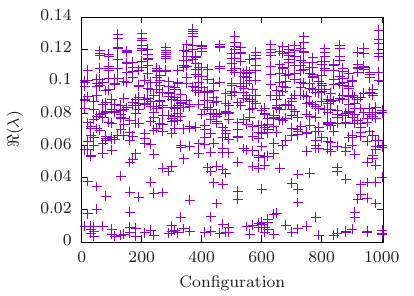}
	\caption{Lowest part of the spectrum of the $\mathcal{O}$ operator at flow time $\tau=0.5$ for 100 configurations.}
	\label{O_spectrum}
\end{figure}

The next parameter that enters in the computation of the AFM is the number of the lowest modes of the $\mathcal{O}$ operator one needs to sum. This is in principle configuration dependent as it depends on the number of instantons of each configuration. However, as we saw in the noisy configuration, we expect that a gap in the spectrum appears and we hope a cut can be chosen consistently for all configurations of a given ensemble. From Fig.~\ref{O_spectrum} we can see that GF increases the gap to the lowest part of the spectrum. For now we will optimize the cut manually such that the AFM reproduces better the topological density from the GF.

Finally we show some characteristic examples for  topological charge densities obtained with the AFM. On a significant subset of the configurations, the AFM and the GF produced compatible topological charge densities. We display in Fig.~\ref{MC_density} some specific configurations which capture two different scenarios for the cases with incompatible densities:

\begin{figure}[h!]
	\centering
	\includegraphics[width=1\linewidth]{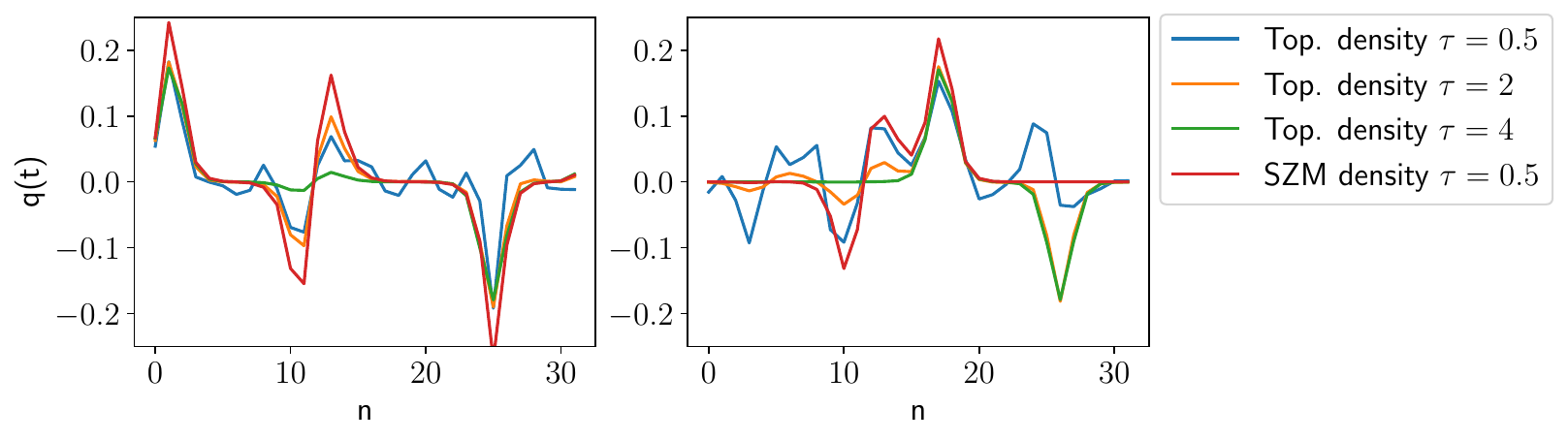}
	\caption{Example configurations comparing the topological density obtained from the AFM at $\tau=0.5$ with the topological density from the gauge field obtained at different flow times.}
	\label{MC_density}
\end{figure}
\begin{itemize}
	\item \textbf{Figure \ref{MC_density} right}: The AFM captures well the topological density at small flow times. An instanton/anti-instanton pair is annihilated in the GF. This shows how the AFM is better at not modifying the underlying gauge configuration.
     \item \textbf{Figure \ref{MC_density} right}: The AFM misses one structure. The index theorem of $D_{ov}$ for this configuration is not fulfilled, possibly due to lattice artifacts and the small size of the spatial 
\end{itemize}

\section{Conclusion}
Our analysis shows that the AFM is capable of filtering the underlying semiclassical content from lattice configurations. Compared to low modes in the fundamental representation, the adjoint representation has the important advantage to reproduce the topological density with only a single lowest eigenmode on smooth configurations. We have tested that the method works well even when only quasi-zero modes are present. We have also shown how the method can filter the noise of several heat bath steps. We noticed that more than one mode is needed in this case to reconstruct the topological charge density. Nevertheless, the required number of modes is still small.

We have applied the method to configurations generated in Monte-Carlo simulations. We obtained promising results for certain configurations where the AFM matched perfectly the results from the GF. However, for some other configurations the method misses some structures and the index theorem is not fulfilled. We believe that the coarseness of the lattice and specifically the very small size of the box are playing a mayor role. The current setup leads to small size instantons on a coarse lattice, which makes it more difficult to distinguish large scale fluctuations from the UV noise. We are currently investigating whether the results are improved towards the continuum limit \cite{Gonzalez-Arroyo:2023}.
A promising aspect of the method is that it reproduces also pairs of instantons and anti-instantons, which tend to be annihilated by the Gradient flow.

\end{document}